\documentclass[12pt,preprint]{aastex}
\usepackage{epsf}

\shorttitle{Quasar Intranight Optical Variability}
\shortauthors{Gopal-Krishna et al. }

\begin{document}

\title{Clear Evidence for Intranight Optical Variability
in Radio-quiet Quasars}

\author{Gopal-Krishna}
\affil{National Centre for Radio Astrophysics, 
Tata Institute of Fundamental Research, 
Pune University Campus,  Post Bag 3, Pune 411007, 
India}
\email{krishna@ncra.tifr.res.in}

\author{C.\ S.\ Stalin\altaffilmark{1}, Ram
Sagar}
\affil{State Observatory, Manora Peak,  Naini Tal, 263129, India}
\email{stalin(sagar)@upso.ernet.in}

\and
\author{Paul J.\ Wiita}
\affil{Department of Physics \& Astronomy, MSC 8R0314,
Georgia State University,
Atlanta, GA 30303-3088}
\email{wiita@chara.gsu.edu}

\altaffiltext{1}{Presently, Visiting Fellow at NCRA/TIFR, Pune}

\begin{abstract}
We present new clues to the problem of the radio loudness
dichotomy arising 
from an extensive search for intranight 
optical variability in seven sets of optically luminous 
radio-quiet quasars and (radio-loud) BL Lacertae objects, 
which are matched in optical luminosity and redshift. 
Our  monitoring of radio-quiet quasars has for the first
time clearly detected such intranight variability,
with peak-to-peak amplitudes $\sim$1\%, occurring with a 
duty cycle of $\sim$ 1/6. The matched BL Lacs have both
higher variability amplitudes and duty cycles when observed in the same
fashion. We show that the much less pronounced intranight 
variability of the radio-quiet quasars relative to  BL Lacs can be 
understood in terms of a modest misalignment of the 
jets in radio-quiet quasars from the 
line-of-sight.   We thus infer that relativistic 
particle jets 
may well also
emerge from radio-quiet quasars,
but while traversing the short
optical-emitting distances, they 
could be snuffed out, possibly 
through inverse Compton losses in the nuclear region. 

\end{abstract}


\section{Introduction}

The dichotomy of radio emission from quasars has been a
persistent hurdle in developing a general theoretical
framework for the emission from active
galactic nuclei (AGN). Whereas 
the powerful jets of relativistic particles are
believed to be generic to the central engines of the radio-loud 
subset 
(e.g.\ Antonucci 1993; Urry \& Padovani 1995), 
the situation 
remains confused as to the existence of such jets
in the radio-quiet majority 
of quasars 
(e.g.\ Antonucci, Barvainis, \& Alloin 1990;
Sopp \& Alexander 1991; Terlevich et al.\ 1992;
Stein 1996; Ivezi{\'c} et al.\ 2002). 

Intranight optical variations (INOV) of blazars, established by using CCDs 
as N-star photometers (e.g.\ Miller, Carini, \& Goodrich 1989;
Carini et al.\ 1992;
Noble et al.\ 1997) are now generally linked to the presence of
relativistic jets (e.g.\ Marscher, Gear, \& Travis 1992;
Wagner \& Witzel 1995; Noble et al.\ 1997). 
Equally clear signatures 
of jets have been lacking for radio-quiet quasars (RQQs), 
despite several
searches for INOV in luminous RQQs 
(Gopal-Krishna, Wiita, \& Altieri 1993; Gopal-Krishna, Sagar, \& 
Wiita 1995; 
de Diego et al.\ 1998; Rabbette et al.\ 1998; 
Gopal-Krishna et al.\ 2000).  
Although radio observations have revealed faint, aligned 
structures in a handful of RQQs
(Miller, Rawlings, \& Saunders 1993; Kellermann et al.\
1994; Papadopoulos et al.\ 1995; Kukula et al.\ 1998; Blundell \& Rawlings 2001)
the case for relativistic jets as a generic feature of RQQs remains 
unsettled (Sopp \& Alexander 1991; Wilson \& Colbert 1995;
Stein 1996;  Kukula et al.\ 1998).

In our earlier papers, statistical 
evidence for intranight optical fluctuations was presented for some
RQQs (Gopal-Krishna et al.\ 1995, 2000; 
Sagar, Gopal-Krishna, \& Wiita 1996),
but in no case was it overwhelmingly convincing.   
The results of several independent 
studies have been discrepant and hence inconclusive
(Jang \& Miller 1995, 1997; de Diego et al.\ 1998; 
Rabbette et al.\ 1998; Romero,
Cellone \& Combi 1999). Jang \& Miller
(1995, 1997) claimed detection of INOV in far more BL Lacs than 
in radio-quiet AGN from a heterogeneous sample.
For the RQQs Ton 951 and Ton 1057, Jang \& Miller (1995, 1997)  presented 
differential light curves (DLCs) showing up to $\sim 8\%$ variations on 
hour-like time scales. However, optical luminosities of both these AGNs 
are modest (M$_B$ $>$ $-$24.3, taking $H_0 = 50$  km s$^{-1}$Mpc$^{-1}$ and 
$q_0 = 0$), and  close to the critical value below which radio 
properties are thought to become like those of Seyfert galaxies
(Miller, Peacock, \& Mead 1990).  At these lower levels of AGN/galactic
light ratios, false indications of variability, produced by seeing 
variations which include different amounts of host galactic light 
within the photometric aperture, 
becomes very probable (Cellone, Romero, \& Combi 2000). 
 Romero et al.\ (1999) monitored a sample of 23 southern objects:
8 RQQs and 15 blazars.  None of their 8 RQQs was clearly found to vary down to 
1\% rms, while 9 of their 15 blazars showed INOV above that level.  
Rabbette et al.\ (1998) also failed to detect INOV in
a sample of 23 high luminosity RQQs, but their
detection threshold was $\sim$ 0.1 mag.
In contrast to these tentative results implying little
INOV for RQQs, de Diego et al.\ (1998) 
concluded that microvariability is
at least as common among RQQs (6 detections in 30 sessions)
as it  is among the (relativistically beamed) 
core-dominated radio-loud quasars (CDQs) (5 in 30), commonly 
deemed as blazars along with BL Lacs.  Each of their 17 RQQs  
had a  CDQ counterpart of
nearly matching brightness and redshift.    
The observational and analysis procedures 
of de Diego et al.\ (1998)
differ  radically  
from those of all other programs, 
including ours. They usually monitored each object only a few (3--9)
times per night at intervals of $\sim$30 minutes; they
divided each of these observations into 5, roughly
one-minute each, exposures.  de Diego et al.\ (1998)  used small
($\sim 2^{\prime \prime}$) apertures and estimated their
errors through an 
analysis of variance technique which involved only 
one comparison star.  Compared to those of other groups,
these techniques lead to less trustworthy results. 


\section{Observations}

Motivated by the need to look for a signature of relativistic 
nuclear jets in intrinsically luminous, {\it bona-fide} RQQs, we 
launched in 1998 a program of R-band monitoring of seven
sets of bright (m$_B$ $\sim$ 16) AGN, each set falling in a narrow 
redshift bin between $z~= 0.17$ and $2.2$, and consisting of a RQQ, 
a BL Lac (except in the highest
$z$ bin), a CDQ and a radio lobe-dominated quasar. 
Thus, the four AGN 
classes in the sample are matched in the $z-M_B$ plane. We
monitored each of the AGN on $\ge~3$ nights, taking $\sim$ 5 
exposures per hour, for durations between 4 and 8 hours per night. 
This program required 113 
nights during 1998--2002, details  of which are
presented elsewhere (Stalin 2002; Gopal-Krishna et al., 
in preparation).
Here we summarize the main 
results obtained for the RQQs and BL Lacs over the course
of 53 nights of observations (Table 1).  All seven 
RQQs are not only optically luminous, ($-24.3 \ge M_B \ge -29.8$)
but also genuinely radio-quiet, with $R < 1$, where $R$ 
is the rest-frame ratio of 5 GHz to 250 nm 
flux densities (Stocke et al.\ 1992).

The R-band CCD observations were made using 
the 1.04-meter
Sampurnanand telescope of the State Observatory, Naini Tal, India.
At least two, but usually more, comparison stars, similar in
brightness to the target AGN were present on each (bias subtracted,
flat-fielded) CCD frame.  We
derived differential light curves (DLCs) of the AGN relative to these
comparison stars and also for all the pairs of comparison stars. Thus, we
identified and discounted any comparison stars which themselves varied.

Photometry of the AGN and the comparison stars was carried out
using the same circular aperture, and the  instrumental magnitudes
were determined using the {\it phot} task in 
IRAF\footnote
{IRAF is distributed by the National Optical Astronomy Observatories which is 
operated by the Association of Universities for research in Astronomy, Inc. 
under co-operative agreement with the National Science Foundation}. 
For each
night a range of aperture radii was considered and the one that minimized
the variance of the DLC of the steadiest comparison star pair was
accepted. The typical aperture radius used was 4$^{\prime \prime}$;
however,
the DLCs are not very sensitive to the chosen radius.
Variations
exceeding 0.01-mag over the night can be readily detected on these DLCs.

\section{Results}
Fig.\ 1 shows the DLCs of two of the RQQs, 1029$+$329 
($R < 0.2$) and 1252$+$020 ($R = 0.5$) (Table 1). 
In each case, the
DLCs of the RQQ against all three comparison stars 
(top three panels) are consistent in showing a gradual fading by $\sim 1\%$ 
over 4--5 hours, whereas the simultaneous 
DLCs involving the same three comparison stars are steady to within $\sim$0.3\%.

Conceivably, the decline in the DLCs of the RQQs could be an artifact of 
the color difference, $\Delta C_{QS}$, between the RQQ and the comparison 
stars, leading to a differential attenuation with varying zenith distance 
(i.e., airmass). 
However, this possibility can be discounted, since no such systematic
fading is evident on the star-star DLCs shown in the two bottom panels
in Fig.\ 1, for which the color difference is comparable to $\Delta C_{QS}$
(except for a brief flare seen near 17.3 UT in the DLC for the star pair 
S2$-$S1, 
which is clearly attributable to a variation of star S2).
Another potential caveat is that a systematic variation in the point 
spread function (PSF) could have led to a varying contribution from the 
RQQ host galaxy within the photometric aperture (Cellone et al.\ 2000). 
This possibility can also be excluded, since
the host galaxy is expected to contribute $ < 10\%$ of the flux 
of each of these luminous RQQs, and is also expected to be encompassed well
within the $\sim 4^{\prime\prime}$ aperture radius used. In addition, 
we have determined
the PSF for the successive CCD frames using the comparison stars
and find that the PSF actually narrowed progressively
by $\simeq 1^{\prime \prime}$ over both of these nights. This 
implies that the actual 
fading of the two RQQs are marginally larger than those recorded on the 
DLCs (Fig.\ 1). We conclude
that the observed INOV of these two RQQs, although small,
is real.  All these checks have not been employed in earlier studies,
so these two cases with well resolved brightness gradients represent the clearest evidence 
reported so far for intranight variability of luminous RQQs. (Similar 
reasoning is applicable to all the cases of INOV reported here). 
The results for the RQQs and  BL Lacs in the 
sample are summarized in Table 1.

There exists a wide discrepancy between the reported duty cycles (DC)
of INOV for RQQs {\it vis-a-vis} BL Lac objects/blazars 
(Jang \& Miller 1997; de Diego et 
al.\ 1998; Romero et al.\ 1999). 
Contributions to the DC are weighted by the number of
hours (in the rest-frame) for which each source was monitored 
(Romero et al.\ 1999),
\begin{equation}
DC = 100 \frac{\sum_{i=1}^{n} N_i(1/\Delta t_i)}{\sum_{i=1}^{n}(1/\Delta t_i)} \ \     \%,
\end{equation}
\noindent where $\Delta t_i = \Delta t_{i,obs}(1 + z)^{-1}$ is the 
duration (corrected for cosmological redshift) of a monitoring session of 
the  source in the selected class; $N_i$ equals 0 or 1, if the object was 
non-variable or variable during $\Delta t_i$, respectively.

For RQQs, counting only the sessions for which INOV was 
positively detected (Table 1), we find that DC = 17\%.
This can be compared with DC = 72\% determined here for the BL Lacs.
Our data also allow, for the first time, estimation 
of DC for different ranges of 
peak-to-peak variability amplitude,  
$\psi \equiv [(D_{\rm max} - D_{\rm min})^2 - 2\sigma^2]^{1/2}$.
 Here, $D$ is the differential magnitude,
$\sigma^2 = \eta^2<\sigma_{\rm err}^2>$, with
$\eta$ the factor by which the average of the measurement errors 
($\sigma_{\rm err}$, as given by ${\it phot}$ algorithm)
should be multiplied; we find $\eta = 1.50$ (Stalin 2002; Gopal-Krishna
et al., in preparation).  The results are given in Fig.\ 2.
All the RQQs have $\psi < 3\%$, and for $\psi < 3\%$, the DCs for BL Lacs 
and RQQs are very similar. However, stronger INOV, with
$\psi > 3\%$, is exclusive to the BL Lacs (DC = 53\%).
Still, we note our sample is small, and it would be very useful to have
similarly sensitive and careful measurements of a larger number of matched pairs
to allow more confident estimates of duty cycles and distributions of $\psi$.

To quantify the variability, we have employed a statistical criterion 
based on the parameter $C$, similar to that
followed by Jang \& Miller (1997), with the added advantage that for each
AGN we have DLCs relative to multiple comparison stars. This allows us to
discard any variability candidates for which the multiple DLCs do not show
clearly correlated trends, both in amplitude and time. We define $C$ for a
given DLC as the ratio of its standard deviation, $\sigma_T$, and the
mean $\sigma$ of its individual data points, $\eta\sigma_{\rm err}$.
This value of $C_i$ for the $i^{th}$ DLC of the AGN has the corresponding
probability, $p_i$, that the DLC is
steady (non-variable), assuming a normal distribution. For a given AGN
we then compute the joint probability, $P$, by multiplying the values of
$p_i$'s for individual DLCs available for the AGN. The effective
$C$ parameter, $C_{\rm eff}$, corresponding to $P$, is given in Table 1 for
each variable AGN; our definition of variability is $C_{\rm eff} > 2.57$, 
which corresponds to a confidence level in excess of 99\%.  
This is followed by the variability amplitude, $\psi$. 
We also note that for these AGN all the DLCs between
comparison stars were found to show statistically insignificant
variability.

\section{Discussion and Conclusions}
To what extent can the observed INOV of 
RQQs be reconciled with the much more pronounced INOV of the 
BL Lacs? Within the canonical jet picture, any 
flux variations associated with the relativistic outflow 
will have their time-scales shortened and amplitudes boosted in the 
observer's frame. As usual, the Doppler factor is 
$\delta = [\Gamma(1 - \beta {\rm cos}(\theta)]^{-1}$, 
where $\beta = v/c$, $\Gamma = (1 - \beta^2)^{-1/2}$ is the bulk
Lorentz factor of the jet, and $\theta$ is its viewing angle.
Then the observed flux, $S_{obs}$ is given in terms of the 
intrinsic flux, $S_{int}$, 
\begin{equation}
S_{obs}  =  \left( \frac{\delta}{1 + z} \right)^{p} S_{int},
\end{equation}
where $p = (2 - \alpha)$ for a continuous jet (e.g.\ Urry \& 
Padovani 1995);
the spectral index $\alpha \equiv d{\rm ln}(S_{\nu})/d{\rm ln}(\nu)$,
and we have assumed 
$\alpha = - 1$ (Stocke et al.\ 1992). 
Similarly, the beaming shortens the observed time 
scale to $\Delta t_{obs} = \Delta t_{int}(1 + z)/\delta$.

The effect of Doppler beaming on the observed DLCs
is illustrated in Fig.\ 3, taking the example of the BL Lac 
object AO 0235$+$164, for which we found a large ($\psi \sim $13\%) and rapid 
($\tau$ $\sim$ 3.5 hr) variation on 1999 November 12. We use the estimated 
$\delta_o$ = 8.1 for this object (Zhang, Fan, \& Cheng 2002)
 to simulate the DLCs for lower values of 
$\delta$, relevant to observers at larger viewing angles.
This mapping is
achieved simply by compressing the observed DLC amplitudes by 
$(\delta/\delta_o)^p$ and, simultaneously, stretching the DLC 
in time by a factor $(\delta_o/\delta)$.  From these 
simulated DLCs 
 it is evident that observers even marginally
misaligned from the jet direction will monitor a drastically reduced INOV,
both in amplitude and rapidity, for the same BL Lac which appears
highly variable to a somewhat better aligned observer (Fig.\ 3). For instance,
if $\theta = 5^{\circ}$ for the jet of AO 0235$+$164, the
estimated $\delta_o$ = 8.1 corresponds to $\beta$ = 0.978. This would
give $\delta \simeq$ 4 and 2, respectively, for modestly misaligned
viewing angles 
of $\theta = 15^{\circ}$ and $\theta = 25^{\circ}$, thought to be typical of 
RQQs (Antonucci 1993; Barthel 1989). 

We thus suggest that the mere low level
of intranight optical variability of RQQs in no way rules out their  
having optical synchrotron jets as active intrinsically as the jets of 
BL Lacs.  The large difference in the radio properties 
could arise from inverse Compton quenching of the 
jet in a majority of quasars, occurring beyond the very small
physical scale 
probed by the nuclear optical synchrotron jet emission. A possible 
signature of such quenching is the hard X-ray spectral tail found in 
some RQQs (George et al.\ 2000).
This emission from the (modestly misaligned) jets is seen despite the extremely strong forward 
flux boosting of the X-rays expected from the inverse Compton scattering 
of external (e.g., broad emission line) photons
by the relativistic jet ($\propto~\delta^{(4-2\alpha)}$, Dermer 1995). 
It remains possible that the weak fluctuations seen in RQQs
arise from a different process, such as fluctuations from an accretion disk
(e.g.\ Mangalam \& Wiita 1993), while the larger ones seen only in
BL Lacs might originate from jets.  Nonetheless,
our observations and analysis lend some support 
to the concept of a jet-disk symbiosis
(e.g.\ Falcke, Malkan \& Biermann 1995)
where jets emerge ubiquitously from accretion flows; hence,
the dichotomy between radio-loud and radio-quiet quasars
need not imply a fundamental difference in their
central engines.

\acknowledgments  
We thank Vasant Kulkarni, John McFarland and Dick Miller for discussions,
Dan Harris and Alan Marscher for
correspondence, and the anonymous referee for helpful suggestions. 
PJW is grateful for support from RPE funds at GSU
and for continuing hospitality at the Department of Astrophysical
Sciences at Princeton.

\clearpage
\begin{table}
\begin{center}
\caption{The sample of radio-quiet quasars and BL Lac objects\tablenotemark{a}\label{tbl-1}}
\baselineskip=20pt
\scriptsize{
\begin{tabular}{lllllllll} 

\tableline
\tableline

   &                 &            &                  &        &          &        &     &    \\ 
Set&  Object         & Other Name &Type              & ~~B    & ~~$M_B$  &  ~~~z  &  N\tablenotemark{b}& 
Observation durations ($h$) and variability status\tablenotemark{c}
               \\

No.&            &                 &                  &        &          &        &   &                      \\ 

\tableline
   &            &                 &                  &        &          &        &   &                      \\
1. & 0945+438   & US 995          &  RQQ             & 16.45  & $-$24.3  & 0.226  & 3 &  8.0(NV), 6.3(NV), 6.6(NV)   \\  
   & 1215+303   & B2 1215+30      &  BL              & 16.07  & $-$24.8  & 0.237  & 4 &  7.0(V,5.5,3.5), 5.9(NV), 5.0(NV), 6.8(V,4.9,1.8)     \\
2. & 0514$-$005 & 1E 0514-0030    &  RQQ             & 16.26  & $-$25.1  & 0.291  & 3 &  5.3(NV), 5.8(NV), 7.5(NV)    \\
   & 1215+303   & B2 1215+30      &  BL              & 16.07  & $-$24.8  & 0.237  & 4\tablenotemark{d} &  7.0(V,5.5,3.5), 5.9(NV), 5.0(NV), 6.8(V,4.9,1.8)      \\
3. & 1252+020   & Q  1252+0200    &  RQQ             & 15.48  & $-$26.2  & 0.345  & 5 &  6.4(V,3.3,2.3), 6.1(NV),4.3(V,3.6,0.9), 4.6(NV), 7.3(NV)  \\
   & 0851+202   & OJ 287          &  BL              & 15.91  & $-$25.5  & 0.306  & 4 &  6.8(V,2.8,2.3), 5.6(V,6.5,3.8), 4.2(V,5.8,5.0), 6.9(V,2.7,2.8)   \\
4. & 1101+319   & TON 52          &  RQQ             & 16.00  & $-$26.2  & 0.440  & 4 &  8.5(NV), 5.6(NV), 6.1(V,2.6,1.2), 5.8(NV)  \\
   & 0735+178   & PKS 0735+17     &  BL              & 16.76  & $-$25.4  &$>$0.424& 4 &  7.8(NV), 7.4(NV), 6.0(NV), 7.3(V,2.8,1.0)  \\
5. & 1029+329   & CSO 50          &  RQQ             & 16.00  & $-$26.7  & 0.560  & 5 &  5.0(NV), 5.3(V,4.3,1.3), 5.8(NV), 8.5(NV), 6.8(V,3.8,1.2)  \\
   & 0219+428   & 3C 66A          &  BL              & 15.71  & $-$26.5  & 0.444  & 7 &  6.5(V,6.0,5.4), 5.7(V,$>$6.6,5.5), 9.1(V,5.8,4.3), 10.1(V,3.5,3.2), \\
   &            &                 &                  &        &          &        &   &  9.0(V,2.9,2.2), 5.1(NV), 5.1(V,$>$6.6,8.0) \\
6. & 0748+294   & QJ 0751+2919    &  RQQ             & 15.00  & $-$29.0  & 0.910  & 6 &  7.6(NV), 8.3(NV), 5.1(NV), 5.4(NV), 6.0(NV), 5.4(NV)\\
   & 0235+164   & AO 0235+164     &  BL              & 16.46  & $-$27.6  & 0.940  & 3 &  6.6(V,$>$6.6,12.8), 6.2(V,3.2,10.3), 7.9(V,2.6,7.6)     \\
7. & 1017+279   & TON 34          &  RQQ             & 16.06  & $-$29.8  & 1.918  & 3 &  7.3(NV), 7.1(NV), 8.1(NV)   \\

\tableline
\end{tabular}

\tablenotetext{a} {Data are taken from V\'{e}ron-Cetty \& V\'{e}ron (1998).} 
\tablenotetext{b} {Number of nights of observation.}
\tablenotetext{c}{NV = not variable, V = variable; when V, followed by $C_{\rm eff}$ 
and $\psi$(\%) values.} 
\tablenotetext{d}{Data taken from the Set 1 which also includes this BL Lac.}

}

\end{center}
\end{table}

\clearpage
\begin{figure}
\vspace*{-5.0cm}
\centerline{\vbox{\epsfxsize=18cm\epsfbox{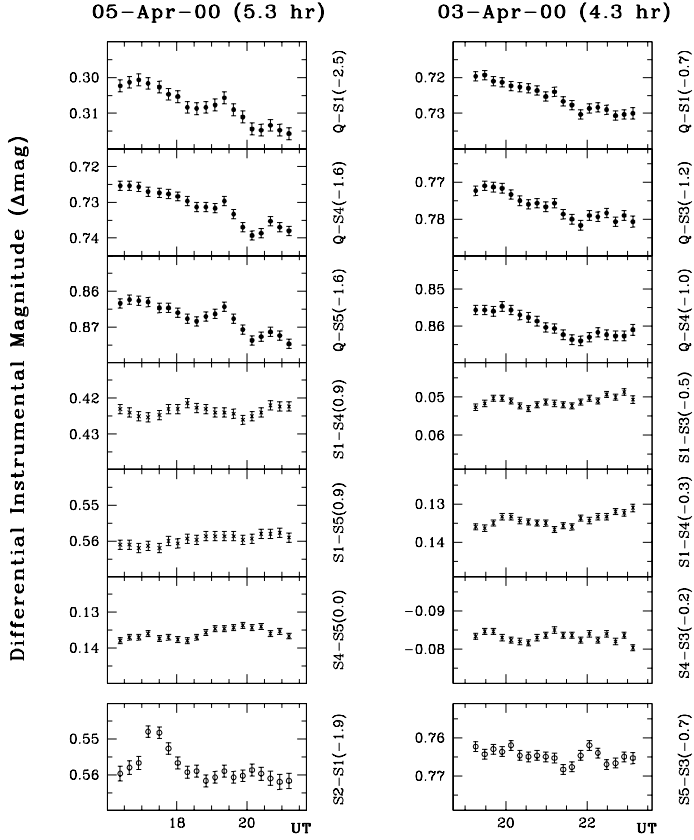}}}
\vspace*{-7.5cm}
\caption{R-band differential light curves (DLCs) for the RQQs 1029$+$329 (left) 
and 1252$+$020 (right), derived using aperture radii of 4$^{\prime\prime}$.1 
and 3$^{\prime\prime}$.6, respectively. 
The top three panels show the DLCs of each RQQ relative to the three comparison 
stars, while the next three panels below display 
the DLCs for the comparison stars, as labeled on the right side. The bottom panel for each RQQ shows 
the DLC for a star pair, also present on the CCD frames, for which the 
differential R$-$B color is comparable to that for the DLCs of the 
corresponding RQQ.  
The J2000 coordinates of the 
stars in the left panels are: 
S1 (10h32m8.94s, $+$32$^\circ$37$^{\prime}$50$^{\prime \prime}$.7),
S2 (31m59.46s, 41$^{\prime}$56$^{\prime \prime}$.1), 
S4 (32m7.50s, 37$^{\prime}$28$^{\prime \prime}$.1), 
and S5 (31m57.24s, 39$^{\prime}$19$^{\prime \prime}$.8).
The corresponding values for the stars in the right panels are:
S1 (12h55m21.00s, +01$^\circ$41$^{\prime}$13$^{\prime \prime}$.9), 
S3 (55m33.90s, 45$^{\prime}$20$^{\prime \prime}$.9), 
S4 (55m15.60s, 43$^{\prime}$54$^{\prime \prime}$.9), 
and S5 (55m36.06s, 42$^{\prime}$4$^{\prime \prime}$.4).
The numbers inside the parentheses 
to the right of the DLCs are the differences between the (R-B) colors 
of the corresponding pair of objects (as taken from the USNO catalog:
http://archive.eso.org/skycat/servers/usnoa).
} 
\end{figure}

\clearpage
\begin{figure}
\plotone{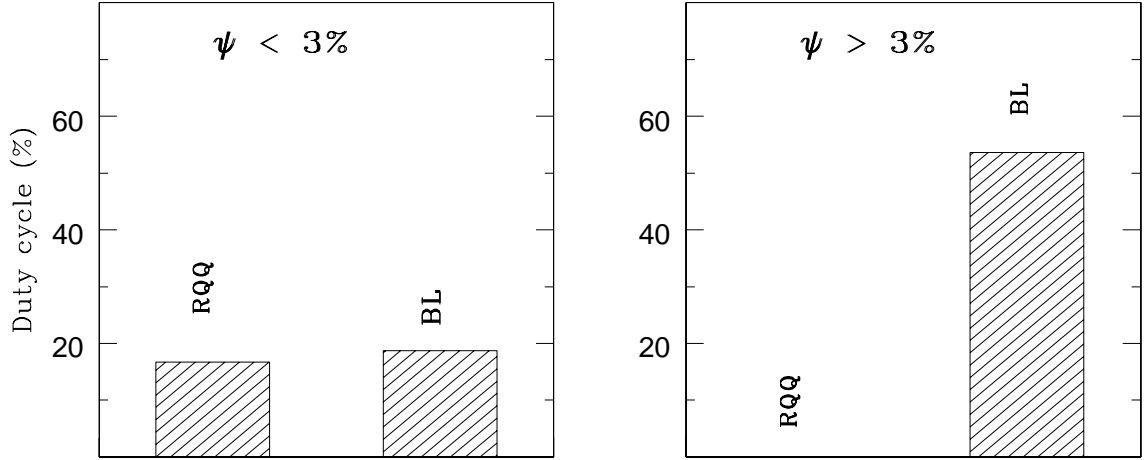}
\caption{Duty cycles of the intranight optical variability (INOV) of the
RQQs and BL Lacs (as determined using the DLCs for all the RQQs and BL Lacs 
in our sample), for two ranges of peak-to-peak variability amplitude, 
$\psi$ (see text). 
The 7 RQQs were observed on 29 nights for a
total of 185.8 hours; the 5 BL Lacs, for 148.1 hours on 22 nights (Table 1).} 
\end{figure}

\clearpage
\begin{figure}
\plotone{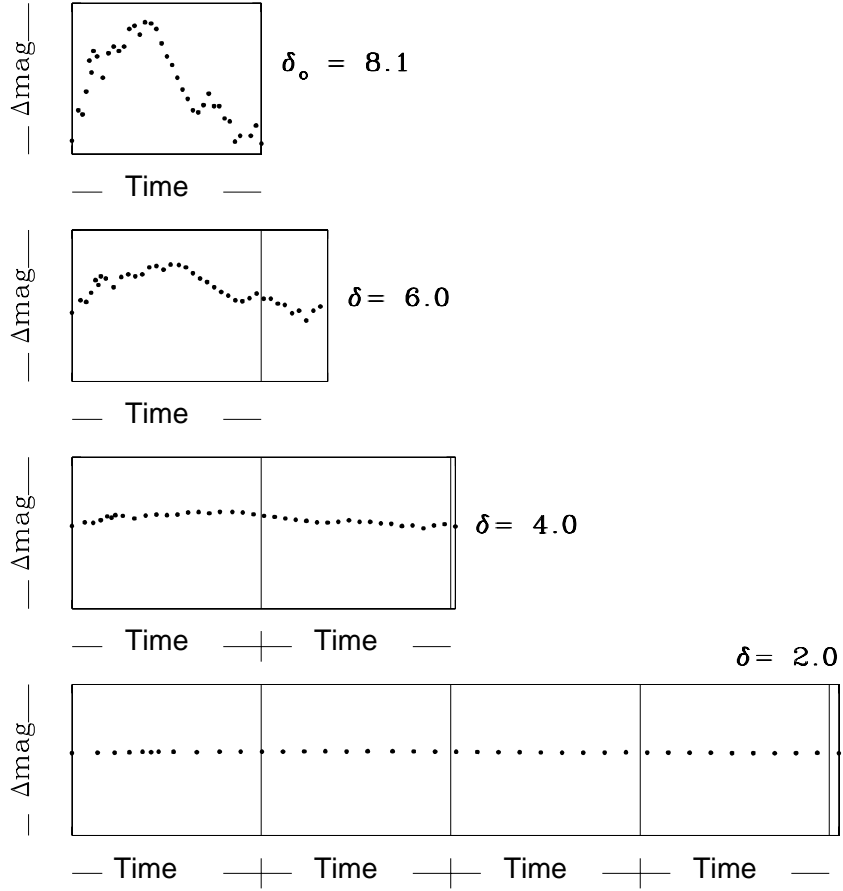}
\caption{The top panel shows the R-band DLC from observations on 
1999 November 12 of the BL Lac object AO 0235$+$164 for which 
$\delta_o = 8.1$
(Zhang et al.\ 2002). The remaining three panels show the DLCs
simulated from the observed DLC, by applying a correction for Doppler
de-beaming appropriate to progressively lower values of $\delta$ 
(which involves an amplitude contraction and temporal stretching,
see text). The total amplitude, $\Delta$ mag, for each panel is 0.1-mag and the 
indicated time duration of each frame in any of the four panels is
6.6 hours (in the observer's frame of reference).} 
\end{figure}

\end{document}